\documentstyle[aps,prl,epsf,preprint]{revtex}
\def\fun#1#2{\lower3.6pt\vbox{\baselineskip0pt\lineskip.9pt
        \ialign{$\mathsurround=0pt#1\hfill##\hfil$\crcr#2\crcr\sim\crcr}}}
\begin{document}
\title{\vskip-2.5truecm{\hfill \baselineskip 14pt {{
\small  FERMILAB-Pub-96/108-A\\
       \hfill CERN-TH/96-131}}\vskip .1truecm} 
\vskip 0.1truecm {\bf Gauge-Mediated Curvature of
the Flat Directions During Preheating
\thanks{Submitted to Phys. Lett. {\bf B}}}}
\author{{Gia Dvali}\thanks{Georgi.Dvali@cern.ch}$^{(1)}$ and 
{Antonio Riotto}\thanks{
 riotto@fnas01.fnal.gov}$^{(2)}$}
\address{$^{(1)}${\it Theory Division, CERN\\Geneva, Switzerland}}
\address{$^{(2)}${\it NASA/Fermilab Astrophysics Center, Fermilab 
National Accelerator Laboratory,\\Batavia, IL 60510, USA}}
\maketitle


\begin{abstract}
\baselineskip 12pt
Usually one expectes the inflaton field to be coupled to some
gauge-charged particles allowing for its decay during reheating.
Such particles then
play a role of the messengers for the gauge-mediated supersymmetry
breaking during and (shortly) after the inflation and radiatively induce soft
masses to all other $D$-flat directions. We show that during the
preheating stage this
gauge-mediated soft masses are typically much greater
than the Hubble parameter during inflation. The dramatic role
is played by the supersymmetry (SUSY) breaking
due to the  parametric resonance  
effect,
which ensures that the inflaton predominantly decays into the bosons
and not the fermions. Difference in the Fermi-Bose
occupation numbers results in the large gauge-mediated soft masses,
which
determine the post-inflationary evolution of the flat directions, suggesting  
that nonthermal phase transitions mediated by gauge messengers may
play a crucial role in the Affelck-Dine  mechanism for the generation of the  
baryon  
asymmetry.
\end{abstract}
\thispagestyle{empty}
\newpage
\pagestyle{plain}
\setcounter{page}{1}
\def\beq{\begin{equation}}
\def\eeq{\end{equation}}
\def\beqa{\begin{eqnarray}}
\def\eeqa{\end{eqnarray}}
\def\tr{{\rm tr}}
\def\x{{\bf x}}
\def\p{{\bf p}}
\def\k{{\bf k}}
\def\z{{\bf z}}
\baselineskip 20pt
In the low-energy minimal supersymmetric standard model there exist a
large number   of $D$-flat directions along which squark, slepton and Higgs
fields get expectation values.  In
flat space at zero temperature
exact supersymmetry guarantees that the effective potential  along
these $D$-flat directions vanishes to all orders in perturbation theory
(besides the possible presence of nonrenormalizable terms in the superpotential
\cite{nr}).
In the commonly studied supergravity scenario, supersymmetry breaking may take
place in isolated hidden sectors \cite{hidden} and then gets transferred to the
other sectors by gravity. The typical curvature of  $D$-flat directions
resulting from this mechanism   is
\begin{equation}
\widetilde{m}^2\sim\frac{\left|F\right|^2}{M_{{\rm Pl}}^2},
\end{equation}
where $F$ is the vacuum expectation value (VEV) of the $F$-term breaking
supersymmetry in the hidden sector. In order to generate soft masses
of order of $M_W$ in the matter sector, $F$ is to be of order of $(M_W\:M_{{\rm
Pl}})$
and sfermion and Higgs masses   along flat directions turn out to be in the TeV
range.
$D$-flat directions do not cause any cosmological problems.  Some of them are
  extremely important in the Affleck-Dine (AD) scenario for   baryogenesis
\cite{ad} if large expectation values along flat vacua are present during  the
early stages of the evolving Universe. This is a necessary
condition
for the AD mechanism to be
operative. In generic supergravity theories 
soft supersymmetric breaking masses are of order of the Hubble parameter  
$H_{{\rm I}}$ (typically $\sim 10^{13}$ or so)
GeV during inflation \cite{dine}. This is due to the fact that
inflation provides a nonzero energy density $V\sim\left|F\right|^2$  which
breaks supersymmetry. Since in the inflationary phase the vacuum energy
dominates, the Hubble parameter is given by
$H^2=(8\pi V/3 M_{{\rm Pl}}^2)$ and therefore the curvature  along the $D$-flat
directions  becomes $\widetilde{m}^2=c H_{{\rm I}}^2$, where $c$ may be either
positive or negative.
This fact has dramatic effects on  what discussed so far. For the AD mechanism,
large squark and slepton VEV's do not result if the induced soft mass squared
is positive, but they do occur if it is negative and an acceptable baryon
asymmetry can be obtained without subsequent entropy releases.

During inflation, $D$-flat directions, however, can get larger gauge-mediated
soft masses \cite{dvali}. It is well known that
in the present vacuum (with zero energy) the gauge interaction can be of
more efficient messenger
of the SUSY breaking than the gravity, provided the messenger scale is
below $M_{{\rm Pl}}$.
This is what usually happens in models with  Gauge-mediated
Supersymmetry Breaking (GMSB) \cite{gaug} where the message about
the supersymmetry breaking (in a gauge-invariant direction) from the hidden
sector  is transferred to the observable sector  through gauge
interactions  by the messenger sector.
The latter is formed by some heavy superfields, transforming under the gauge  
group $G$ as a real or conjugate  representation, which suffer from a tree
level supersymmetry breaking. 
The crucial point, however, is that even if gauge-mediated
corrections are zero in the present vacuum
they had to be important during inflation if the
superfield $X$, which is dominating  
inflation, is coupled to some of the gauge nonsinglet superfields $\phi$
\begin{equation}
W = g\:X\:\phi^2.
\end{equation}
In most of the inflationary scenarios such couplings are expected
to be there, in order to allow for the efficient reheating through the
final decay of the inflaton field.
In such a case $\phi$-fields would play a role of the messengers of the
gauge-mediated SUSY breaking during inflation and all other
$D$-flat directions would obtain a radiative two-loop mass
given by  \cite{dvali}
\begin{equation}
\widetilde{m}^2\sim
\left(\frac{\alpha}{4\:\pi}\right)^2\:g^2\:\frac{\left|F_X\right|^2}{M_\phi^2},
\end{equation}
where $M_\phi\sim g\:\left|X\right|$ is the mass term of the $\phi$, $\alpha$  
is the gauge coupling of the gauge group $G$ and $g$ is the coupling constant
relating the superfield $\phi$ to the inflaton. Using again the relation  
between the vacuum energy and the Hubble constant during inflation
$H_{{\rm I}}^2\sim \left|F_X\right|^2/M_{{\rm Pl}}^2$,
one can rewrite the above  
relation as
\begin{equation}
\label{two}
\widetilde{m}^2\sim H_{{\rm I}}^2\:
\left(\frac{\alpha}{4\:\pi}\right)^2\:\left(\frac{M_{{\rm
Pl}}}{\left|X\right|}\right)^2,
\end{equation}
which shows that, in general, $\widetilde{m}^2$ may be larger than
$H_{{\rm I}}^2$ (typical magnitude of the soft masses induced by gravitational
sources) if $\left|X\right|$ is somewhat
below $M_{{\rm Pl}}$ \cite{dvali}. Notice that these corrections are
independent of coupling constant $g$.
Although there is no generic proof, usually, e.g. for the low representations
of the simple Grand Unified Groups (GUT's),  these two-loop radiative  
corrections to
 $\widetilde{m}^2$ are positive  
and the $D$-flat directions are expected to be stabilised
at the origin during
inflation. This would be a disaster for the AD mechanism of baryogenesis  
which requires  
large expectation values along the flat vacua after inflation to be operative. 
However, it has been recently point out \cite{dvalinew} that if the inflaton
couples to the superfields of the messenger sector and the latter are in the
complex representation, supersymmetry breaking during inflation can generate
one-loop Fayet-Iliopoulos $D$-terms. The corresponding soft masses are
proportional to the abelian generators of $G$ ({\it e.g.} hypercharge in the
GUT's) and, therefore, can have either sign. They
can dominate the gauge-mediated two-loop soft breaking terms, being of the
order of
\begin{equation}
\label{nnn}
\widetilde{m}^2\sim H_{{\rm
I}}^2\:\left(\frac{\alpha}{4\:\pi}\right)\:\frac{M_{{\rm pl}}}{\left|X\right|}.
\end{equation}
Such (negative) masses can destabilize the
 sfermion flat directions during inflation, playing 
a crucial role for the AD  mechanism of baryogenesis. 
Previously the  induced gauge-mediated soft masses where
analysed only during inflation.
However, the crucial role for the post-inflationary
evolution of the flat directions is played by their soft masses
just before the reheating process.  
The aim of this letter is to analyse the issue of gauge-mediated
supersymmetry breaking
during a particular stage of the evolution of the early universe. 
The epoch we are referring to is
called preheating   \cite{explosive} and is expected to occur after the end    
of chaotic
inflation. The crucial relevance of supersymmetry breaking at preheating has
been
first pointed out in \cite{linderiotto}.
At the very beginning of this period, which is dominated by the coherent
oscillations of the inflaton field, one can distinguish two possible
cases depending whether the  classical expectation value of the
messenger\footnote{Below we will refer
to the gauge-charged superfields coupled to
the inflaton as `messengers', although they are not assumed to
be necessarily a messengers in the present vacuum, but only in the
early universe.} is  fixed at its minimum or undergoes  coherent
oscillations together with the inflaton. The latter will be the case if
the VEV of some $\phi$ component is nonzero in the minimum
about which the inflaton oscillates.
We will start considerations from the former case assuming no coherent
oscillations of $\phi$.
Kofman,  Linde  and  Starobinsky  have recently pointed out  that
the explosive decay of the inflaton occurs at the first stage of reheating
through the phenomenon of parametric resonance \cite{explosive}. The inflaton
energy
is released in the form of inflaton decay
products, whose occupation number is extremely large, and have
energies much smaller than the temperature that would have been
obtained by an instantaneous conversion of the inflaton energy density
into radiation.
Since it requires several scattering times for the low-energy decay
products to form a thermal distribution, it is rather reasonable to
consider the period in which most of the energy density of the
Universe was in the form of the nonthermal   quanta produced by
inflaton decay as a separate cosmological era, dubbed as preheating to
distinguish it from the subsequent  stages  of particle decay and
thermalization which can be described by the techniques developed in  
\cite{tec}.
Several aspects of the theory of explosive reheating have been studied in the
case of slow-roll inflation \cite{noneq} and first-order inflation
\cite{kolb}.
One of the most important consequences of the stage of preheating is the
possibility of  nonthermal phase transitions with symmetry restoration
\cite{KLSSR,tkachev,rt}.
These phase transitions appear due to extremely strong quantum corrections
induced by particles produced at the stage of preheating.
What is crucial for our considerations is that
 parametric resonance is a phenomenon peculiar of particles obeying
Bose-Einstein statistics.   Parametric resonant decay into fermions is very
inefficient because of Pauli's exclusion principle. This means that during
the preheating period the Universe is only populated by a huge number of
  soft bosons and the  occupation numbers of bosons and fermions
belonging to the supermultiplet coupled  to the inflaton superfield are
completely unbalanced \cite{linderiotto}. Supersymmetry is then strongly broken during the
preheating era \cite{linderiotto} and large loop corrections may arise since  
the usual
cancellation between diagrams involving bosons and fermions within the same
supermultiplet
is no longer operative \cite{linderiotto}.  We shall see that  the curvature along $D$-flat
directions
during the preheating era  is much larger than the effective mass
 that they  acquire in
the inflationary stage. This makes the details
of the effective potential
along $D$-flat directions  during inflation almost irrelevant as far the
initial conditions
of the condensates along the $D$-flat directions is concerned.
Let us first assume that the  $F_X$-term corresponding to  the superfield $X$
is dominating inflation and that the gauge-charged superfield $\phi$
is in the real (say adjoint) representation of $G$. 
The simplest superpotential
(leading to chaotic inflation \cite{linde} and to the subsequent
resonance decay of the inflaton) one can envisage relating $X$ to  
the supermultiplet $\phi$  is
\begin{equation}
W = M_X\:X\:Z + g\:X\:\phi^2
\end{equation}
where $Z$ is another gauge singlet superfield\footnote{Without
the $Z$ superfield the preheating is necessarily marked by the coherent
oscillations of $\phi$ (see the text below), in which case the parametric
resonance requires further investigation. Here we want to make situation 
maximally adequate to the one studied in \cite{explosive}.} and
$\sim 10^{13}$ GeV for the density perturbations generated during
the inflationary era to be consistent with COBE data \cite{cobe}.  
$G$-invariant contraction of the indices is assumed. There are several
possible choices of the discrete or continuous symmetries under which the
above form is the most general renormalizable one.
One example is a phase symmetry
under which $X \rightarrow e^{i\theta}X$ and
$Z \rightarrow e^{-i\theta}Z$ and
$\phi \rightarrow e^{-i\theta/2}\phi$.
The global minimum of the theory is at $X = 0$ and 
$\phi = \sqrt{-MZ/g} = {\rm arbitrary}$.
For any non-zero value of $X$ the minimum in all other fields is
at  $Z = \phi = 0$ and their masses are $M_Z^2 = M_X^2$ and
$M_{\phi}^2 = g^2|X|^2$ respectively. Therefore, assuming the chaotic
initial conditions $|X| \gg M_X$, we expect that
$\phi$ -field will quickly settle at the origin due to very large
curvature in its direction. Contrastly, the curvature in the
$X$-direction is small and
inflation occurs during the slow rolling of the scalar field $X$ from its
very large value\footnote{The curvature in the  $Z$-direction is also small, so
that in principle both singlets can roll slowly, however for simplicity we
assume that inflation in the $X$ direction lasts longer, so that when it
starts oscillations about the minimum, $Z$-field is already fixed there.}.
Then inflaton oscillates with an initial amplitude $X_0\sim
10^{-1}\:M_{{\rm Pl}}$. 
Within few dozen oscillations the initial energy density 
$\rho_X\sim M_X^2\:X_0^2$ is transferred through the interaction
$g^2\:X^2\:\phi^2$ to  {\it bosonic} $\phi$-quanta
in the regime of parametric resonance \cite{explosive}. At the end of the broad
parametric resonance the field $X$ drops down to $X_e\sim 10^{-2}\:M_{{\rm  
Pl}}$ and parametric resonance only occurs if $g X_e> M_X$. This implies  
$g>10^{-4}$. 
Notice that the flatness of the inflaton  potential during  
inflation  is preserved for such large values of couplings $g$ by  
supersymmetric cancellations. In the above example the one-loop
corrections to the inflaton potential are simply zero, because no
Fermi-Bose mass splitting occurs along the inflationary trajectory
(the only non-zero
$F$-component is the one of the
$Z$-superfield $F_Z = M_XX$, which does not couple
to the other fields). Below we will consider the case when the inflation
is dominated by $F_X$ and one-loop corrections are present, but, in any case,
they only modify the inflaton potential by a logarithmic factor with
a small coefficient).
 At the end of the preheating
era the Universe is   expected to be   filled up with noninteracting
$\phi$-bosons with
relatively small energy per particle, $E_\phi \sim 10^{-1} \sqrt{g\:M_X 
M_{\rm Pl}}$
and  with very large occupation numbers $n_\phi/E_\phi^3 \sim g^{-2}$. 
Here we are assuming that the energy $E_\phi$ is larger than
any bare mass  of
the superfield $\phi$ (which is automatically the case in the above model).
Our results do not crucially  
depend upon this assumption. 
The leading
contribution to the curvature of $D$-flat directions 
comes  from the two-loop exchange of the $\phi$-bosons which are produced
during the  parametric decay of the inflaton  and form the noninteracting gas
of particles out of equilibrium during the preheating stage. Unfortunately, one
cannot use the standard imaginary-time formalism  since in the
nonequilibrium case there is no
relation between the density matrix of the system and the time
evolution operator, which is of essential importance in the
formalism. There is, however, the real-time formalism of Thermo Field
Dynamics (TFD), which suites our purposes \cite{tfd}. This approach leads to a
$2\times 2$ matrix structure for the free propagator for the $\phi$-boson (only
the (11)-component is physical)
\begin{eqnarray}
\left(\begin{array}{cc}
D_{11}(K) & D_{12}(K)\\
D_{21}(K) & D_{22}(K)
\end{array}\right)&=&
\left(\begin{array}{cc}
\Delta(K) & 0\\
0 & \Delta^*(K)
\end{array}
\right)
+\left(\begin{array}{cc}
f_\phi(k) & \theta(k_0)+f_\phi(k)\\
\theta(-k_0)+f_\phi(k) & f_\phi(k)
\end{array}
\right)\nonumber\\
&\times &\:2\pi \delta[K^2-m_\phi^2],
\end{eqnarray}
with the usual vacuum Feynman propagator
\begin{equation}
\Delta(K)=\frac{i}{K^2-m_\phi^2+i\epsilon}.
\end{equation}
The distribution function $f_\phi$ is chosen such that
the number density of particles, $n_\phi=(2\pi)^{-3}\int d^3 \!p \, f_\phi(p)$
and setting it equal to $\sim \rho_\phi/E_\phi$. Notice that at the preheating
stage the occupation number $f_{\tilde{\phi}}$ of the fermionic partner
$\tilde{\phi}$ of the $\phi$-boson is much smaller than $f_\phi$:
even though supersymmetric cancellation may occur when only vacuum propagators
are inserted, such a cancellation is no longer operative in the gas of
$\phi$-bosons where
$f_\phi\gg f_{\tilde{\phi}}$.
Making use of  the standard  TFD Feynman rules one can show that during the
preheating era $D$-flat directions acquire a correction to the mass squared
\begin{equation}
\label{nn}
\widetilde{m}^2\sim  \alpha^2\:\frac{n_\phi}{E_\phi}\sim
\frac{10^{-2}}{g}\:\alpha^2\: M_X\:M_{{\rm Pl}},
\end{equation}
which is much larger than the two-loop contribution $\sim (\alpha/4\:\pi)
M_X^2$ that soft breaking
terms may receive during inflation. 
Now let us consider the case when the messenger field undergoes the
coherent oscillations driven by the oscillations of the inflaton.
This will happen when the instant
VEV of $\phi$ is a nontrivial function of the inflaton VEV.
Such a behaviour is exhibited already by a simplest system: single inflaton
superfield coupled to the messengers
\begin{equation}
W = {1\over 2} \left( M_X\:X^2  + g\:X\:\phi^2\right).
\end{equation}
The global minimum is at $X = \phi = 0$, but for  
$0 < |X| < X_c = {M_X \over g}$ the instant minimum of $\phi$ is at
$|\phi| = \sqrt{{2 \over g}|X|(M_X - g|X|)}$. Thus, whenever $X$ drops
below $X_c$, $\phi$ will undergo the driven
coherent oscillations. For $X > X_c$, $\phi$ vanishes and the
tree level potential is dominated by the inflaton $F$-term
$F_X = M_XX$, which
splits masses of the Fermi-Bose components in the $\phi$
superfield. This splitting result in two things:
1) the one-loop corrections
to the inflaton slope, which for large $|X|$ behave as
\begin{equation}
\left (\Delta V_{{\rm eff}} \right )_{|X|\rightarrow \infty}
\sim  {g^2 \over 16\pi^2}\:M_X^2\:|X|^2\:{\rm ln}|X|^2;
\end{equation}
and 2) the two-loop universal (up to charges) gauge-mediated soft
masses for the $D$-flat directions
\begin{equation}
\widetilde{m}^2 \sim \left({\alpha \over \pi}\right)^2\:M_X^2
\end{equation}
After the inflaton VEV drops below the critical value $X_c$ both
fields start to oscillate about the global minimum.
Parametric resonance in such a case needs a special investigation,
which will not be attempted here. Instead we will argue that there is
an independent source of the supersymmetry breaking due to a coherent
oscillations of the $\phi$ VEV. This condensate can be regarded
as a gas of cold bosons with energies $\sim M_X$ and occupation numbers 
$n_{\phi} \sim {M_X^3 \over g^2}$. Again, since there are no fermions
the two-loop gauge diagrams do not cancel out and the resulting soft
masses can be estimated as
\begin{equation}
\widetilde{m}^2 \sim {\alpha^2 \over g^2}\:M_X^2
\end{equation} 
Although these masses are smaller than (9), they are greater than
the gravity-mediated contribution and thus, will play a dominant role
in the cases in which parametric resonance is suppressed.
Let us now assume another case, as  suggested in \cite{dvalinew},   
that the inflaton couples to some superfields of the messenger sector
belonging
to the complex representations.
We introduce a pair of messengers $\phi$ and $\bar{\phi}$ with an opposite
charges under a certain $U(1)$-group. We will think of this $U(1)$
as being an abelian subgroup of some Grand Unified Theory
symmetry under which
$\phi$ and $\bar{\phi}$  transform in the complex representations.
The simplest superpotential which leads to the messenger VEVs being
fixed in their (minimum all the way until the inflaton settles in the
global vacuum) has the form
\begin{equation}
W = W_0 + g\:X\:\bar{\phi}\:\phi
\end{equation}
where $W_X$ is a part of the superpotential responsible for the
slope of inflaton potential, which can be taken to be
$W_0 = M_X\:X\:Z$ as in (6).
The inflation in this model will proceed in the same way as discussed above,
except the preheating stage. The crucial difference is that now inflaton
through the parametric resonance will decay into two different bosons 
$\phi$ and $\bar{\phi}$ which in general can have different occupation
numbers. This can be the case if, for instance, one of these particles
has a nonzero bare mass because of mixing with some other superfield $A$
in the superpotential\footnote{The mass of the $\phi$-quanta induced by  
$D$-terms in presence of a nonvanishing  VEV during inflation along AD flat  
vacua is zero.}
\begin{equation}
 W' = M_A\:\phi \:A
\end{equation}
and the bare mass $M_A$ is so high to stop the production of $\phi$-quanta  
during the parametric resonance. However, if the scale $M_A$ is very high
(much greater than $\sqrt{M_X\:X_e}$), it will suppress the production of  
$\bar{\phi}$
quanta as well. This is because of the superdecoupling arguments: below
the energies $\sim M_A$ the $\phi$ and $A$ fields decouple and the
low-energy superpotential can not include any gauge-invariant coupling
of $X$ and $\bar{\phi}$ superfields. However, these arguments
are not applicable if the SUSY-breaking scale during oscillations
(in our case $\sim \sqrt{M_X\:X_e}$) is comparable to $M_A$. So with
$M_X\sim 10^{13}$ GeV the right order of magnitude for $M_A$ would
be somewhere around the GUT scale.
 Another possible source of the asymmetry between $\phi$ and $\bar{\phi}$
states can be their different cross couplings with an inflaton field
in the K\"ahler potential
\begin{equation}
C^2 \:\int d^4\theta \:{1 \over 4\:M^2} XX^+\bar{\phi}\bar{\phi}^+ =
C^2\: {|F_X|^2 \over M^2}|\bar{\phi}|^2 + ...
\end{equation}
Such couplings with $M \sim {M_{{\rm Pl}} \over \sqrt{8\pi}}$ will
generically be presented in supergravity theories. Assuming $g = 0$
and $W_0= M_X\:X^2$ in (14), this interaction induces an effective
cross coupling $|X|^2|\bar{\phi}|^2$ in the potential with the coefficient
$g_{{\rm eff}}^2 \sim C^2 \left  ({M_X \over M} \right )^2$
\footnote{In the previous example, 
the coupling (16) with $M = M_A$ can be generated for very large values of
$M_A$ after integrating out the heavy $\phi$ and $A$ states. However, in
this case, as a matter of the potential structure,
one can not use simply $W_0 = M_X\:X^2$.}.
Then,
the initial energy density of the inflaton 
$\rho_X\sim M_X^2\:X_e^2$ may be  transferred through the interaction
$g_{{\rm eff}}^2\:X^2\:|\bar{\phi}|^2$ to  {\it bosonic}
$\bar{\phi}$-quanta
in the regime of parametric resonance \cite{explosive}.
 Because of the mass difference, at the preheating stage the
messengers $\phi$ and $\bar{\phi}$  have
different number densities, $n_{\bar{\phi}}\gg n_\phi$.  Their
contributions to the curvature along $D$-flat directions do not cancel in the
one-loop diagrams
giving rise to a Fayet-Iliopoulos $D$-terms. The corresponding soft masses
for sfermion fields are proportional to
\begin{equation}
\widetilde{m}^2\sim\left(4\:\pi\alpha\right)\:  
\frac{n_{\bar{\phi}}}{E_{\bar{\phi}}} 
\sim \left(4\:\pi\alpha\right)\:
 \frac{10^{-2}}{g_{{\rm eff}}}\: M_X\:M_{{\rm Pl}},
\end{equation}
which is larger than the one-loop correction (5) obtained by soft  
breaking masses during inflation.
Our result implies  that during the explosive stage of preheating
gauge-mediated supersymmetry  
breaking is stronger than at the stage of  inflation,  
suggesting that nonthermal phase transitions mediated by gauge messengers may  
play a crucial role in the AD mechanism for the generation of the baryon  
asymmetry \cite{linderiotto}. 

\acknowledgments
AR would like to thank Greg Anderson and Andrei Linde for many useful
discussions. AR is
supported by
the DOE and NASA under Grant NAG5--2788.
\begin{enumerate}

\end{enumerate}

\end{document}